# Remarkable enhancement in catechol sensing by the decoration of selective transition metals in biphenylene sheet: A systematic first-principles study

*Vikram Mahamiya[a,†], Juhee Dewangan [a,†], Alok Shukla[a], Brahmananda Chakraborty[b,c]*

[a]Indian Institute of Technology, 400076 Mumbai, India

[b]High Pressure and Synchrotron Radiation Physics Division, Bhabha Atomic Research Centre, Bombay, Mumbai 40085, India

[c]Homi Bhabha National Institute, Mumbai 400094, India

Email:mahamiyavikram@gmail.com, shukla@phy.iitb.ac.in, brahma@barc.gov.in

## Abstract

Motivated by the recent successful synthesis of biphenylene structure [*Science* 372, (2021), 852], we have explored the sensing properties of this material towards the catechol biomolecule by performing the first-principles density functional theory and molecular dynamics simulations. Pristine biphenylene sheet adsorbs catechol molecule with a binding energy of -0.35 eV, which can be systematically improved by decorating the transition metals (Ag, Au, Pd, and Ti) at various possible sites of biphenylene. It is observed that the catechol molecule is adsorbed on Pd and Ti-decorated biphenylene sheets with strong adsorption energies of -1.00 eV and -2.54 eV, respectively. The interaction of the catechol molecule with biphenylene and metal-decorated biphenylene is due to the charge transfer from the O-2p orbitals of the catechol molecule to the C-2p orbitals of biphenylene and d-orbitals of metals in metal-decorated biphenylene, respectively. From the Bader charge calculation, we found that 0.05e amount of charge is transferred from the catechol molecule to pristine biphenylene, which gets almost double (~0.1e) for the Ti-decorated biphenylene sheet. The diffusion energy barrier for the clustering of the Pd and Ti atoms comes out to be 2.39 eV and 4.29 eV, computed by performing the climbing-image nudged elastic band calculations. We found that the catechol molecule gets desorbed from the pristine biphenylene sheet even at 100 K but remains attached to metal (Pd, Ti) decorated biphenylene sheets at room temperature by performing the *ab-initio* molecular dynamics simulations. The Ti-decorated biphenylene sheet has more sensitivity toward catechol adsorption while the Pd-decorated biphenylene sheet has a suitable recovery time at 500 K. The results suggest that the Pd and Ti-decorated biphenylene sheets are promising materials for catechol detection.

# 1  Introduction

The combustion of fossil fuels through industrial and transport activities produces many toxic species harmful to the environment and human health. The gas molecules such as $NO_2$ [1], NO [2], $NH_3$ [3], CO [4], etc., are toxic and cause detrimental environmental effects. The detection of toxic materials is essential for medical diagnosis, detecting ecological pollutants, food monitoring, and so forth [5]. Effective sensing of these poisonous species is challenging for the scientific community. The sensing substrate should have good sensitivity, selectivity, fast recognition, and reliability [6]. The catechol molecule also called pyrocatechol [7], is a phenolic substance [8] with the molecular formula of $C_6H_4(OH)_2$ [7]. It is used to prepare pharmaceuticals, cosmetics, insecticides, antioxidants, etc. [7]. Since catechol is a toxic and low degradable molecule, its unchecked exposure to the environment for a long time is harmful and may negatively affect the environment and human health. Catechol may govern several health problems like skin irritation, eye damage, genetic defects, and so forth [9]. So, the advancement of sensing materials with high sensitivity, and fast recovery time, has become a strict necessity for catechol recognition.

There are many theoretical and experimental approaches reported for successful catechol detection. For instance, Nasim et al. [8] introduced a novel enzyme-based biosensor using an artificial neural network successfully tested in natural water samples for catechol detection. They found that the catechol detection limit of 0.032 μM. Wang et al. [10] investigated a reduced graphite oxide incorporated into a metal-organic framework MIL-101 (Cr) that is highly sensitive and reliable with the catechol detection limit of 4 μM. Yang et al. [11] explored the sensing behavior of borophene for catechol compound using DFT methods and obtained a suitable and concise recovery time of 7.6 ns. The sensing performance of ZnO/RGO nanocomposites toward catechol is investigated theoretically and experimentally [9]. The ZnO/RGO composites exhibit excellent sensitivity of 162.04 μA mM $cm^{-2}$ and a lower detection limit of 47 nM [9]. Using DFT simulation, the catechol detection behavior of metal-decorated two-dimensional dichalcogenide $MoS_2$ is studied by Lakshmy et al. [7]. It has been found that the Ti-decorated $MoS_2$ system is a potential material for catechol sensing [7]. Manjunatha et al. [12] fabricated a poly modified graphene electrode for catechol detection in a water sample with a lower detection limit of $8.7\times10^{-7}$ mol $L^{-1}$. It has been reported that the boron-doped carbon quantum dots can detect the catechol molecule by an on-off fluorescent switching method with an ultrasensitive detection limit [13]. Qian et al. [14] fabricated an electrochemical electrode-based high-performance polymeric film sensor with an excellent catechol detection performance. Zheng et al. [15] obtained the N-doped carbon nanotube electrodes by heat treatment with the precursor of ZIF-67, which exhibits outstanding performance towards catechol sensing with a recovery range of 95.3-105.0%. Zhao et al. [16] prepared the nickel oxide deposited carbon nanotube sensor for catechol recognition, revealing remarkable sensitivity with a detection limit of 2.5 μM. Bansal et al.[17] studied the synthesis and characterization of zirconia nanoparticles for catechol monitoring on water samples with an exquisite sensitivity of 0.14 μA/μM $cm^2$. Balakumar et al.[18] have synthesized an Au@NG-PPy nanocomposite for catechol determination, which shows excellent electrocatalytic activity towards catechol with a detection limit of 0.0016 μM. Yan et al.[19] fabricated catechol photo electrochemical sensors, which offer a wide detectable range and low detection limit toward catechol. Kumar et al. [20] investigated the electrochemical sensing of catechol, and they found excellent sensitivity and acceptable recovery with a low detection limit of 0.95 μM for catechol detection. It has been studied that the nanoporous gold thin films show a superb detection range for catechol [21].

Generally, the orbital interaction and charge transfer between the analyte and the host material is small, which is difficult to find out in experiments. By employing first-principles theoretical calculations, one can observe the change in the density of states, band structures, etc., and examine the charge transfer, which is one of the most crucial aspects of electrochemical sensing. From the theoretical point of view, the charge transfer makes some changes to the properties of the host material, such as resistivity, optical or electronic band gap, etc., which generates a signal to be detected by the sensor [5,25]. The signal that is generated may be optical or chemical, according to which the sensor is designed. If the optical band gap, absorption, conductivity, etc., of the host material is changed after the signal generation, then it is an optical sensor [5]. One can also find suitable metals for the decoration purpose to enhance the sensing ability of the host material by theoretical modeling, which can help the experimentalist perform experiments.

Two-dimensional (2D) nanomaterials, including graphene [26], graphyne [27], porous nanomaterials [28–30], transition metal dichalcogenides [31,32], etc., have gained a lot of scientific interest in sensing applications due to their large surface area, excellent electrochemical and optical properties. These properties are highly tunable by applying strain or decorating metal atoms at various possible sites, which make them a more promising substrate for hazardous gas or biomolecule sensing [5]. The specificity of the 2D materials is high sensitivity and selectivity towards different biomolecules, lightweight, high charge mobility, long-term reusability, and portability, which are essential features, especially for sensing applications [5]. In addition, these porous nanomaterials provide various channels for gas diffusion that can reduce the recovery time of the sensor. Biphenylene (BP) is one of the novel two-dimensional (2D) graphene-like carbon allotrope synthesized by Schlutter et al. in 2014 [33]. The BP sheet comprises square, hexagon, and octagon carbon rings [34], connected with $sp^2$ and $sp^3$ hybridization bonds. In the structure of BP, a large number of possible sites are presented for the decoration of metal atoms so that we can get the desired electrochemical properties [2,35]. Because of the presence of a large number of adsorption sites in the BP structure, it possesses outstanding behavior in various fields such as energy storage [36], gas sensing [2], Li-ion batteries [37], and so forth [38]. Recently, a theoretical study on catechol sensing has been performed in a similar type of 2D nanomaterial holey graphyne (hGY) decorated with an Sc atom [22]. This study reports the suitable catechol adsorption energy of -3.22 eV and a total amount of 0.9e charge transfer from the catechol molecule to the host structure. BP sheet also contains six and eight-membered rings for the metal decoration; therefore, it will be interesting to study the catechol sensing properties of metal decorated BP monolayer.

In the present study, first-principles density functional theory and *ab-initio* molecular dynamics simulations have been performed to explore the catechol sensing behavior of the pristine and metal decorated BP sheets. We have considered Ag, Au, Pd, and Ti transition metal atoms (TMs) for the decoration purpose. We found that the Pd and Ti atoms are strongly bonded to the BP sheet with higher binding energies than Ag and Au. Hence, we have considered only Pd and Ti atom decoration on the BP sheet for catechol adsorption. Since the charge transfer mechanism is the central point to describe the sensing behavior of the substrate, we have performed the partial density of states, Bader charge analysis [39], and charge density difference calculations. Ti-decorated BP sheet has more sensitivity towards the catechol detection due to the more charge transfer from the catechol molecule to the metal atom than the Pd decorated BP sheet. In addition to that, the recovery time is checked for the different temperatures at different frequencies. We found that the recovery time for the BP + Pd + catechol system is 0.90 seconds and 47 milliseconds, respectively, for the yellow and UV lights at 500 K, which is very suitable for practical applications. The results demonstrate that the Pd and Ti-decorated BP systems have promising potential in catechol sensing because of their excellent sensing attributes. The movement of a metal atom, thermal stability, and metal-metal clustering issues are treated by the diffusion energy barrier and AIMD calculation.

## 2  Computational details

We have used the density functional theory and *ab-initio* molecular dynamics methods as incorporated in Vienna Ab Initio Simulation Package (VASP) [40]. The Projector Augmented Plane wave (PAW) method within Generalized Gradient Approximation of Perdew-Burke-Ernzerhof (GGA-PBE) [41] exchange-correlation functional is adopted, and the energy cut-off is chosen to be 500 eV. The Hellman-Feynman force and energy convergence limit are set as 0.01 eV/Å and $10^{-5}$ eV, respectively. The Monkhorst-Pack k-point grid of 5×5×1 and 7×7×1 points are taken for geometry optimization and density of states calculations, respectively. A 2×2×1 supercell of BP containing 24 carbon atoms is considered, and the vacuum of 30 Å is taken perpendicular to the sheet to avoid interaction between periodic layers. Grimme's dispersion correction DFT-D3 [34, 35] is applied with GGA-PBE to include weak van der Waals (vdW) interactions. The *ab-initio* molecular dynamic simulations (AIMD) [44] are performed to investigate the thermal stability of the system at room temperature. The system is placed in the microcanonical ensemble (NVE) followed by the canonical ensemble (NVT) for five ps time durations.

## 3  Results and discussion

### 3.1  Catechol detection in pristine BP sheet

We have considered a 2×2×1 supercell of BP containing 24 carbon atoms for all the calculations. The structure of BP comprises 4, 6, and 8 atoms of carbon rings, as presented in **Fig. 1(a)**. The optimized lattice constants of the BP unit cell are a = 4.52 Å, and b = 3.75 Å, matching excellently with the reported literature values of 4.52 Å, and 3.76 Å [45]. We have obtained four different C-C bond lengths in the BP sheet, which have values of 1.452 Å, 1.446 Å, 1.459 Å, and 1.405 Å, consistent with the literature [38]. The unit cell of BP is presented in **Fig. 1(b)**.

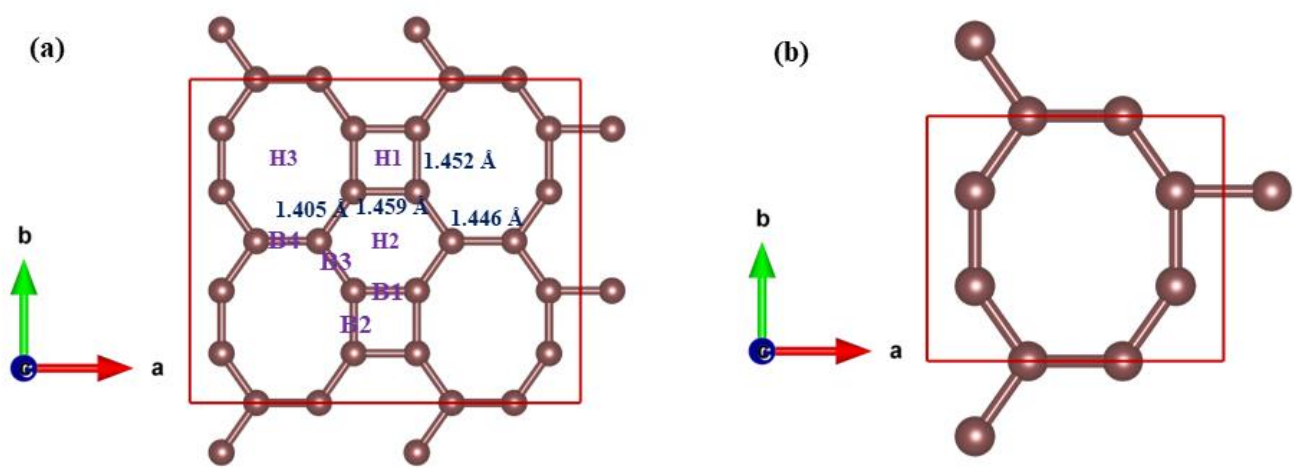

**Fig. 1 Optimized structure of (a) 2*2*1 supercell of BP with different carbon–carbon bond lengths and possible adsorption sites. (b) unit cell of BP. Here brown colour balls represent carbon atoms.**

To optimize the catechol molecule, we have placed it at the center of a cubic box with a lattice dimension of 20 Å, and geometry relaxation calculations are performed. The optimized structure of the catechol biomolecule is presented in **Fig. 2(a)**.

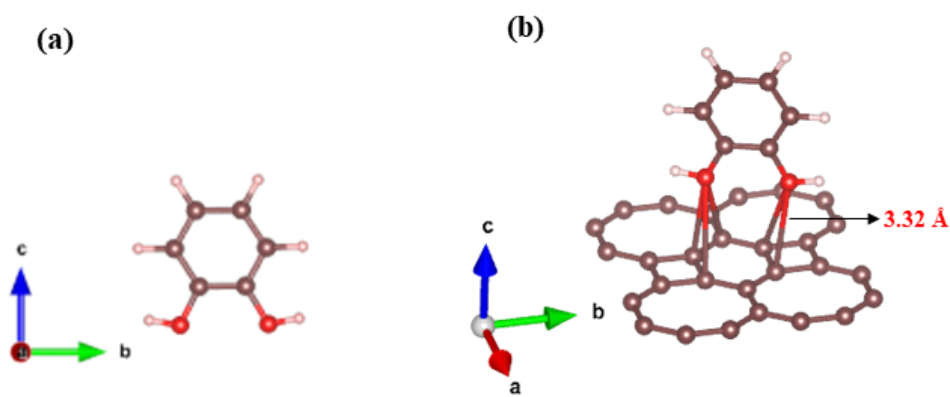

**Fig. 2** Optimized structure of (a) catechol molecule (b) catechol molecule attached to BP sheet at H2 site. Here brown, red and pink colour balls represent carbon, oxygen and hydrogen atoms, respectively.

Here, we have performed DFT calculations within GGA approximations and DFT-D3 vdW corrections. People have also performed adsorption energy calculations of dopants by applying DFT along with the local density approximation (LDA) and B3LYP functionals [11,46,47].

After optimizing the pristine BP and catechol molecule structure, the catechol molecule is placed at a 2 Å distance above the BP sheet at possible adsorption sites, and system is relaxed. The adsorption energy of the catechol molecule is calculated by using the following equation,

$$E_{ads}(catechol) = E(BP + catechol) - E(BP) - E(catechol) \quad (1)$$

where, $E(BP + catechol)$ is the total energy of the BP + catechol system, $E(BP)$ is the total energy of pristine BP sheet, and $E(catechol)$ is the total energy of the catechol molecule. The maximum adsorption energy of the catechol molecule when it is placed above the hexagon is found to be -0.35 eV, with C–O bond length of 3.32 Å. The relaxed structure of the BP + catechol system is presented in **Fig. 2(b)**.

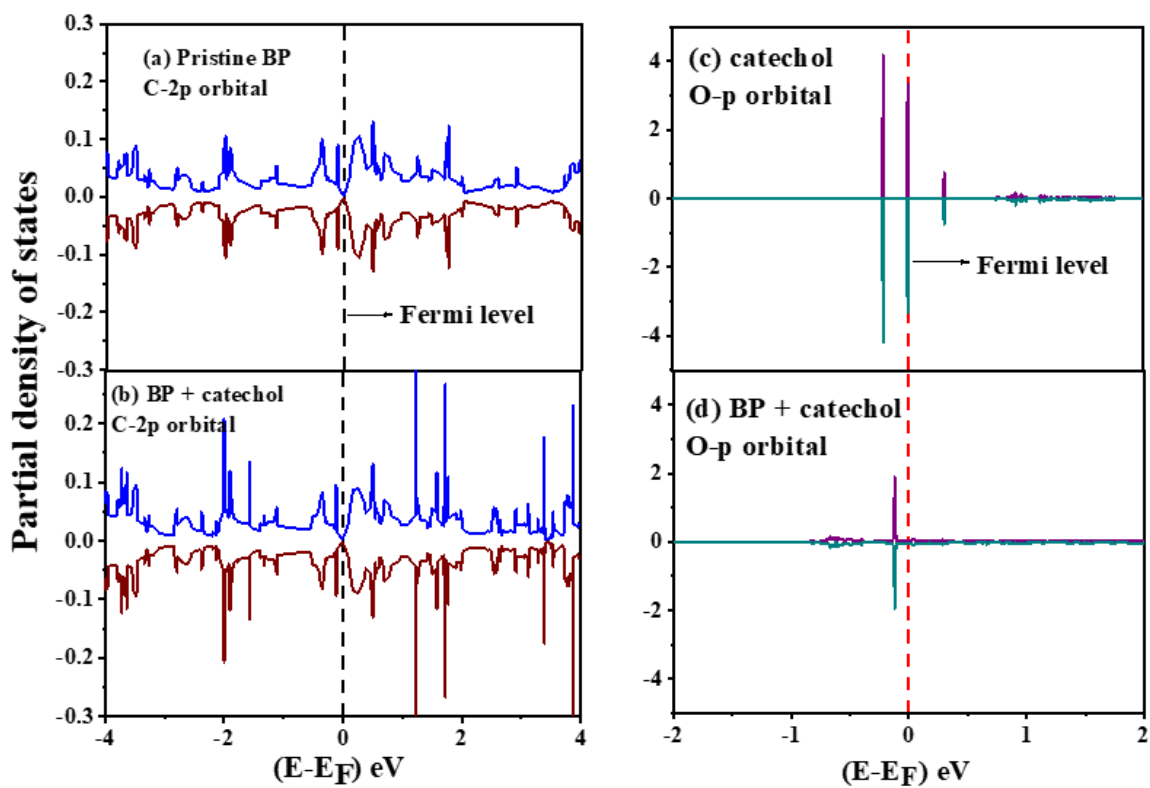

**Fig. 3** Partial density of states plots for (a) C-2p orbitals of pristine BP sheet (b) C-2p orbitals of BP + catechol system (c) O-2p orbitals of catechol molecule (d) O-2p orbitals of BP + catechol system. The Fermi level is set to zero eV.

The catechol adsorption energy for pristine $VSe_2$ is reported to be -0.25 eV by Chakraborty et al. [46], and for pristine hGY, it is -0.70 eV reported by Lakshmy et al. [22]. We have performed the partial density of states (PDOS) and Bader charge [39] calculations to study the charge transfer mechanism. The PDOS plots for C-2p orbitals and O-2p orbitals before and after the adsorption of the catechol molecule on a pristine BP sheet are presented in **Fig. 3**. When the catechol molecule is attached to the BP sheet, the states for the O-2p orbitals of the catechol molecule are depleted near the Fermi level (**Fig. 3(d)**), which signifies the charge loss from the O-2p orbital. Thus, the charge is transferred from the O-2p orbitals of the catechol molecule to the C-2p orbitals of the pristine BP sheet. From the Bader charge calculations, we found that a very less amount of charge 0.05e is going from the catechol biomolecule to the BP sheet. Hence, the adsorption of the catechol molecule on pristine BP is very weak, and the pristine BP may not be suitable for catechol detection.

### 3.2 Adsorption of transition metals (TMs) on pristine BP sheet

The adsorption energy of adsorbed biomolecule/gas on the host material is linked to the sensitivity of the sensing device such that the higher the adsorption energy, the better the sensitivity of the sensor [5]. Hence to elevate the sensitivity of pristine BP towards catechol, a suitable approach is to decorate the BP sheet with TMs. The TMs, including Ag, Au, Pd, and Ti having valence electron configuration $4d^{10} 5s^1$, $5d^{10} 6s^1$, $4d^{10}$, and $3d^2 4s^2$, respectively, are considered to decorate the BP sheet in this work. We have selected TMs of different d-series and having 1 and 2 s-valence electrons for the thorough investigation of charge transfer. Although the metal (Ag/Au/Pd/Ti) decorated BP systems have not been fabricated, several noble metal-decorated nanomaterials are experimentally synthesized as an effective and low-cost sensor [48–50]. Metal decoration can be experimentally challenging and costly, however, in our proposed sensor the loading of metal atom is very small, so the designed sensor should be affordable for large scale synthesis. The possible adsorption sites for the decoration of TMs are H1, H2, H3, B1, B2, B3, and B4, as depicted in **Fig. 1(a)**. To calculate the binding energy of the metal atom, the TMs are placed one by one at around a 2Å distance above the all-possible adsorption sites, and each system is optimized. The optimized structures of Ti-decorated on the BP sheet at various adsorption sites are presented in **Fig. S1 (a, b, c, d, e, f, & g)**. The following equation calculates the binding energy of the TM atoms on the the BP sheet:

$$E_b(TM) = E(BP + TM) - E(BP) - E(TM) \tag{2}$$

where, $E(BP + TM)$ is the energy of the TM-decorated BP sheet, $E(BP)$ is the energy of the pristine BP sheet, and $E(TM)$ is the energy of the TM atom. The comparative study of binding energies of all the TMs, including their positions prior to and after the decoration on the BP sheet, is summarized in **Table 1**.

**Table 1. Various metal decorated BP systems with position of metal atom before and after its adsorption on BP sheet, binding energy and C–metal bond length. The notation H1, H2, H3, B1, B2, B3, and B4 are defined in Fig. 1(b) in manuscript.**

| System | Initial position | Final position | Binding energy (eV) | C–metal bond length (Å) |
|---|---|---|---|---|
| Biphenylene + Ag | H1 | H1 | -0.29 | 3.41 |
| Biphenylene + Ag | H2 | H2 | -0.31 | 3.65 |
| Biphenylene + Ag | H3 | H3 | -0.55 | 2.77 |
| Biphenylene + Ag | B1 | B1 | -0.29 | 2.56 |
| Biphenylene + Ag | B2 | Moved slightly towards H3 | -0.54 | 2.52 |
| Biphenylene + Ag | B3 | Moved slightly towards H3 | -0.54 | 2.43 |
| Biphenylene + Au | H1 | B2 | -0.85 | 2.37 |
| Biphenylene + Au | H2 | H2 | -0.60 | 3.72 |
| Biphenylene + Au | H3 | H3 | -1.18 | 2.17 |
| Biphenylene + Au | B1 | B1 | -0.82 | 2.37 |
| Biphenylene + Au | B2 | Moved slightly towards H3 | -1.18 | 2.17 |
| Biphenylene + Au | B3 | Moved slightly towards H3 | -1.18 | 2.17 |
| Biphenylene + Pd | H1 | H1 | -1.90 | 2.27 |
| Biphenylene + Pd | H2 | H2 | -1.44 | 2.39 |
| Biphenylene + Pd | H3 | H1 | -1.90 | 2.27 |
| Biphenylene + Pd | B1 | H1 | -1.90 | 2.27 |
| Biphenylene + Pd | B2 | H1 | -1.90 | 2.27 |
| Biphenylene + Pd | B3 | B3 | -1.77 | 2.15 |
| Biphenylene + Pd | B4 | B3 | -1.77 | 2.15 |
| Biphenylene + Ti | H1 | H1 | -2.79 | 2.17 |
| Biphenylene + Ti | H2 | H2 | -3.67 | 2.17 |
| Biphenylene + Ti | H3 | H3 | -2.99 | 2.24 |
| Biphenylene + Ti | B1 | B1 | -3.85 | 2.09 |
| Biphenylene + Ti | B2 | B1 | -3.85 | 2.09 |
| Biphenylene + Ti | B3 | B1 | -3.85 | 2.09 |
| Biphenylene + Ti | B4 | B4 | -2.18 | 2.07 |

We found that the maximum binding energy for Ag atom is -0.55 eV at the H3 adsorption site, and for Au, it is -1.18 eV at the H3 site. When the Pd atom is placed above the H3 site, it moves to the H1 site after the relaxation, and the obtained maximum binding energy is -1.90 eV. In the case of the Ti atom, the maximum binding energy obtained is -3.85 eV at the B1 site, and at the H2 site, the binding energy is -3.67 eV, which is slightly smaller than the maximum binding energy. The binding energy of the Ag and Au atoms attached to the BP sheet is small, so we have not considered Ag and Au decorated BP sheets for the catechol adsorption.

To understand the electronic properties of pristine and metal decorated BP systems, we have plotted the total density of states (TDOS) for pristine BP, BP + Pd, and BP + Ti systems as presented in **Fig. S2 (a, b, & c)**. From the TDOS plot, the BP is metallic, and when metal atoms are attached to a BP sheet, the states near the Fermi level slightly increase. Hence, metal decoration alters the electronic property of the BP sheet which is a crucial feature for sensing applications. The spin-up and spin-down channels are symmetric for BP + Pd system while asymmetric for the BP + Ti system; hence the structure became magnetic in the case of Ti-decoration. This is due to the presence of the unpaired d-electrons in the valence sector of the Ti atom.

To get a qualitative picture of charge transfer and understand the interaction between TM atoms and the pristine BP sheet, we have calculated the partial density of states (PDOS) for C-2p orbitals and Ti-3d orbitals before and after the adsorption of the Ti atom to the pristine BP sheet, as shown in **Fig. 4 (a, b, c, & d)**. When the Ti atom is attached to pristine BP, the states of the C-2p orbital get enhanced near the Fermi level, which can be seen in **Fig. 4 (b)**, signifying the charge gain by C-2p orbitals of the BP sheet. To get a more precise insight into charge transfer, we have plotted the PDOS for Ti-3d orbital for isolated Ti atom and BP + Ti system, as presented in **Fig. 4 (c & d)**. For an isolated Ti atom, there is an intense state, as shown in **Fig. 4 (c)**, and the intensity of states decreases when the Ti atom is attached to BP.

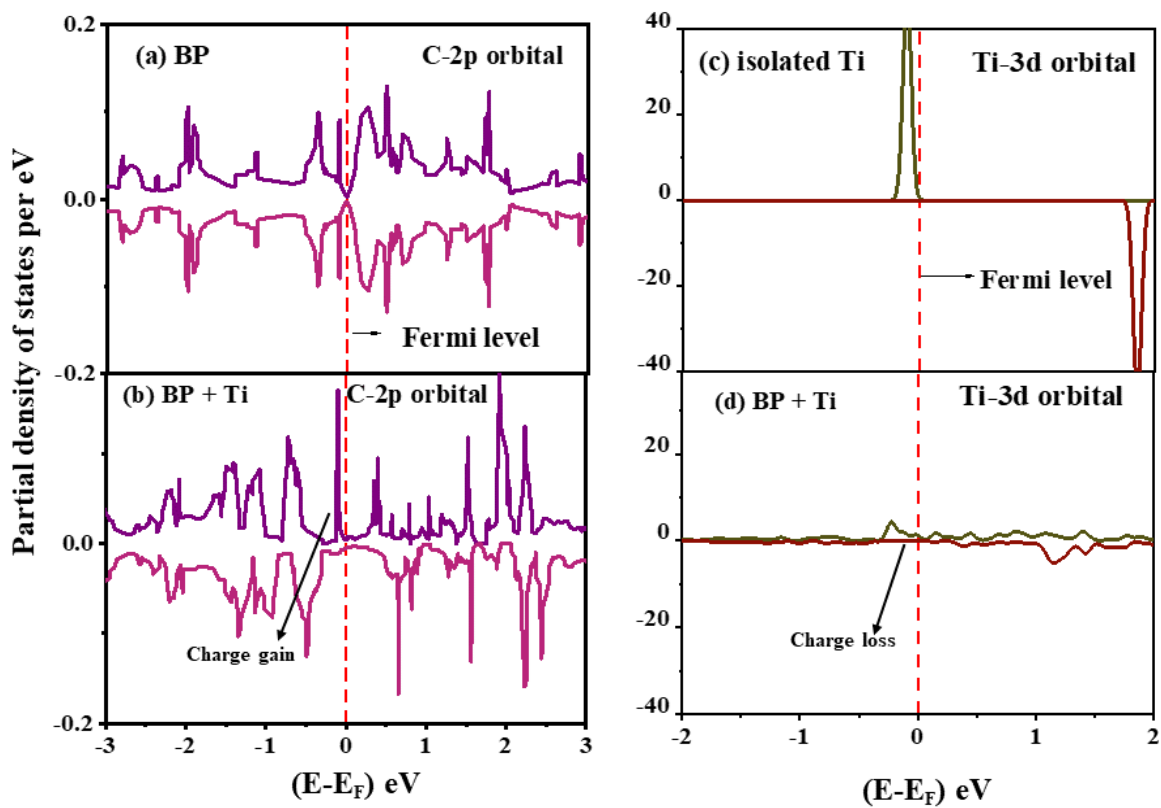

**Fig. 4** Partial density of states plots for (a) C-2p orbital of pristine BP (b) C-2p orbital of BP + Ti system (c) Ti-3d orbital of isolated Ti (d) Ti-3d orbital of BP + Ti system. The Fermi level is set to zero eV.

The decrease in intensity of states signifies that the Ti atom is losing some charge when it is attached to BP. Hence the charge is transferred from Ti-3d orbitals to C-2p orbitals, and this charge transfer is the reason behind the strong binding of the Ti atom with the BP. We have also plotted the PDOS for the C-2p orbitals and Pd-4d orbitals before and after the adsorption of the Pd atom on the BP sheet, presented in **Fig. S3 (a, b, c, & d)**. From the PDOS plots, we can observe that when the Pd atom is attached to the BP sheet, some charge is transferred to the C-2p orbitals from Pd-4d orbitals.

After getting the qualitative idea of the charge transfer, we have performed the Bader charge calculations [39] to get the quantitative idea of charge transfer. We report that a total amount of 1.19e charge is transferred from Ti-3d orbitals to C-2p orbitals when the Ti atom is attached to the BP sheet, while a small amount of charge 0.27e is transferred from the Pd-4d orbitals to the C-2p orbitals of the BP sheet.

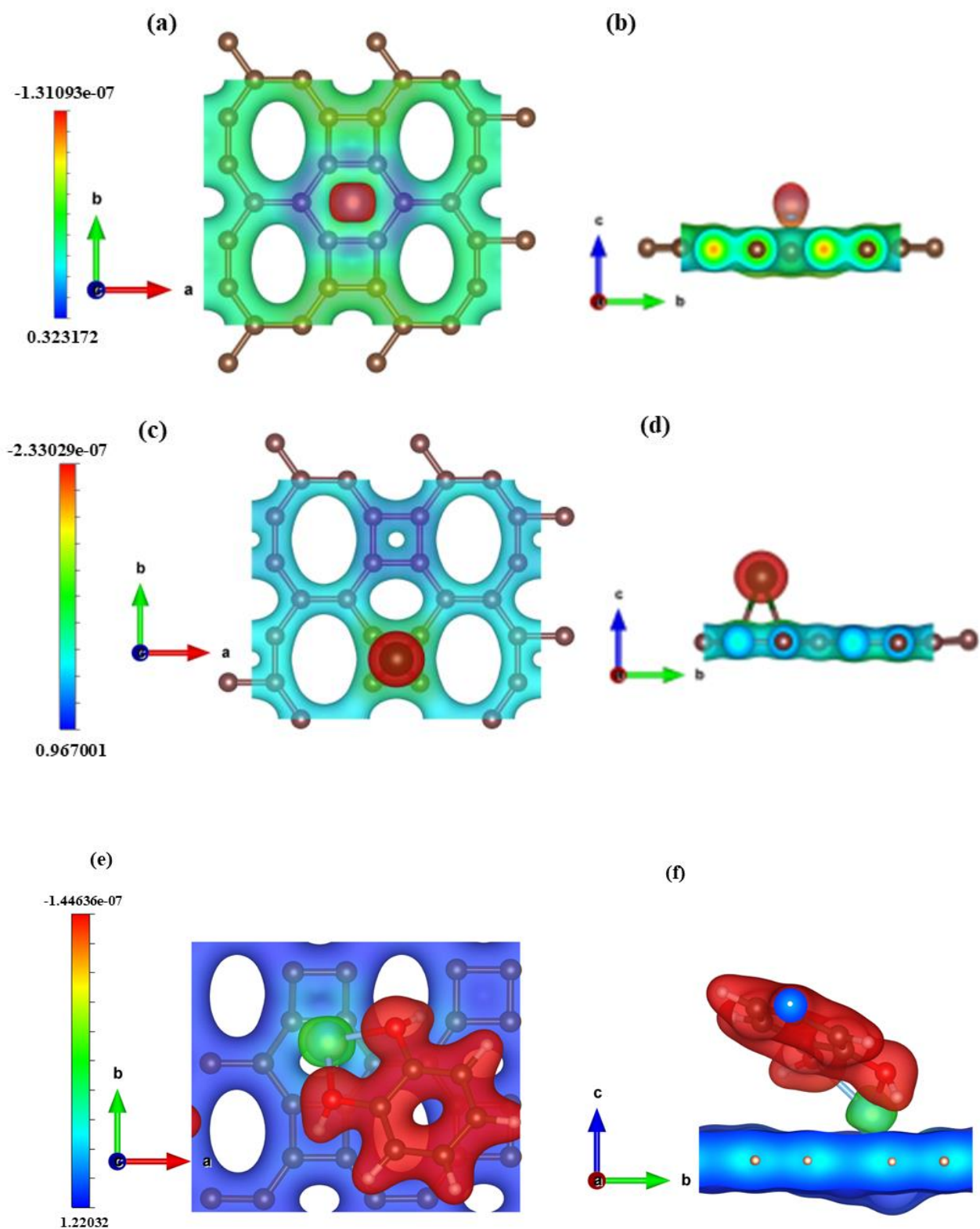

Fig. 5 Charge density difference plots between Ti-decorated BP and pristine BP for isosurface value of 0.07e (a) front view (b) side view. Charge density difference plot between Pd-decorated BP and pristine BP for isosurface value of 0.134e (c) front view (d) side view. Charge density difference plot between BP + Ti + catechol, BP + Ti, and isolated BP for the isosurface value of 0.060e (e) front view, and (f) side view. The BGR colour pattern is adopted for plotting. Here the red colour represents the charge loss region and green or blue colour represent the charge gain regions.

We have drawn the charge density difference plot to visualize the charge transfer process when the Ti atom is attached to the BP sheet, as displayed in **Fig. 5(a & b)**. The charge density difference is plotted between Ti-decorated BP and pristine BP system, i.e., $\Delta\rho = \rho_{BP+Ti} - \rho_{BP}$ for isovalue 0.07e. The B-G-R colour pattern is adopted to plot the charge density difference. In the plot, the red colour region near the Ti atom indicates the charge loss region, and the green and blue colour regions in the BP sheet indicate the charge gain regions. Hence from the charge density difference plot, the charge is transferred from Ti-3d orbitals to C-2p orbitals. The charge density difference between BP + Pd and BP, i.e., $\Delta\rho = \rho_{BP+Pd} - \rho_{BP}$ is presented in **Fig.5 (c & d)** for isovalue of 0.134e. In this plot, the red colour region near the Pd atom represents charge loss region, and the green and blue color regions in the BP sheet represent the charge gain regions.

### 3.3 Catechol detection in TM-decorated BP sheet

We have discussed in the previous section that we will consider only BP + Pd and BP + Ti systems for the catechol detection. To investigate the adsorption of the catechol molecule in the TM-decorated BP system, the catechol molecule is placed around 2 Å above the TM atom, and the relaxation calculations are performed. The geometry optimized structures of BP + Pd + catechol and BP + Ti + catechol is presented in **Fig. 6(a & b)**.

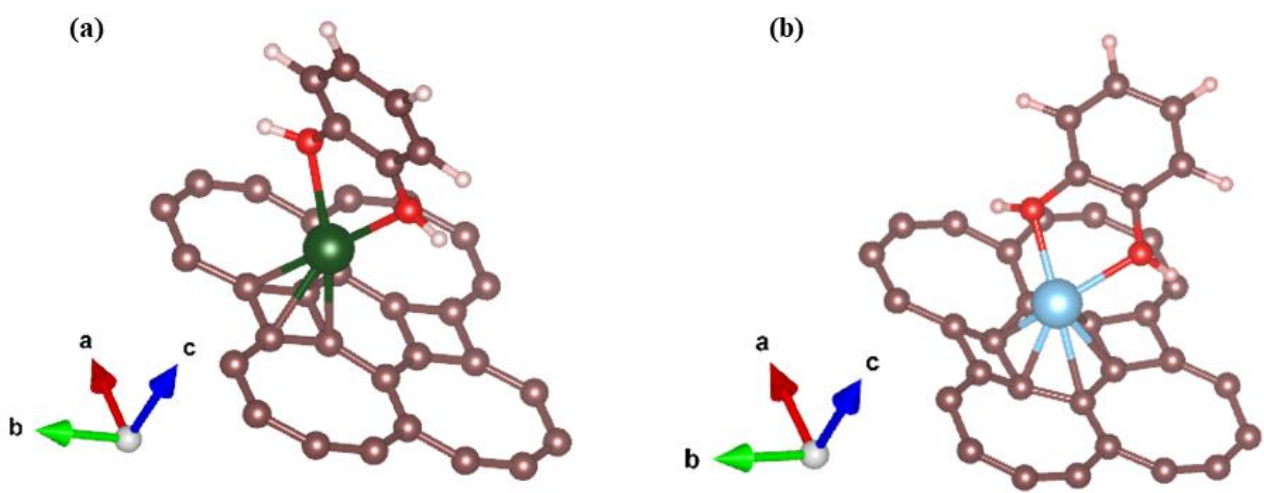

**Fig. 6 Optimized structure of catechol molecule adsorbed on metal-decorated BP (a) BP + Pd + catechol (b) BP + Ti + catechol. Here brown, green, blue, red and pink colour balls represent C, Pd, Ti, O and H atoms, respectively.**

The following equation estimates the adsorption energy of the catechol molecule attached to metal decorated BP sheets,

$$E_{ads}(catechol) = E(BP + TM + catechol) - E(BP + TM) - E(catechol) \quad (3)$$

where, $E(BP + TM + catechol)$ is the energy of BP + Ti/Pd + catechol system, $E(BP + TM)$ is the energy of Ti/Pd-decorated BP system, and $E(catechol)$ is the energy of the catechol molecule. The catechol adsorption energies for BP + Pd and BP + Ti systems are obtained to be -1.00 eV and -2.54 eV, respectively. The Pd–O and Ti–O bond distances are 2.42 Å and 2.21 Å, respectively. Chakraborty et al. [46] have reported the catechol adsorption energy of -0.95 eV for Pd-decorated $VSe_2$. The catechol adsorption energy for Sc-decorated hGY is reported to be -3.22 eV by Lakshmy et al. [22] using the first-principles DFT method. Ponnusamy et al. [51] investigated the RGO-supported ZnO system experimentally and theoretically and obtained the catechol adsorption of -4.09 eV. Lakshmy et al. [7] reported the catechol adsorption energy of -1.79 and -2.23 eV for the $MoS_2$ + Pd and $MoS_2$ + Ti systems, respectively.

The changes in electronic structure properties in pristine BP and metal-decorated BP after the catechol adsorption are determined by the TDOS plots, which is presented in **Fig. S4 (a, b & c)**. For both the configurations, the states near the Fermi level are enhanced, which signifies that the catechol molecule attached to the BP + Pd and BP + Ti systems. Up and down spin states are symmetric for the BP + Pd + catechol system, whereas the spin states are asymmetric for BP + Ti + catechol system; hence the latter possess the magnetic attributes. The magnetic behaviour is due to unpaired 3d electrons in the Ti atom. The catechol adsorption energies and bond lengths on pristine and TM-decorated BP sheets are summarized in **Table 2**.

**Table 2. Catechol adsorption energy, nearest bond lengths between catechol and BP/TM-decorated BP sheets, and charge transfer from catechol to host material.**

| Composition | Adsorption energy (eV) | Bond length (Å) | Charge transfer |
|---|---|---|---|
| BP + catechol | -0.35 | 3.32 | 0.05e |
| BP + Pd + catechol | -1.00 | 2.41 | 0.15e |
| BP + Ti + catechol | -2.54 | 2.24 | 0.10e |

The charge transfer mechanism is the central point of the sensing behaviour. So, to get a clear picture of charge transfer after the catechol adsorption on BP + Ti system, we have investigated the PDOS of outer shell orbitals. The PDOS is plotted for Ti-3d orbitals and O-2p orbitals before and after the adsorption of the catechol molecule to the Ti-decorated BP sheet and presented in **Fig. 7**. From **Fig. 7 (b)**, we can see the enhancement in states near the Fermi level compared to **Fig. 7 (a)**, which signify the transfer of charge to Ti-3d orbitals when the catechol molecule is attached to BP + Ti system. To get a clearer insight into charge transfer, we further calculated the PDOS for the O-2p orbitals of catechol, as presented in **Fig. 7 (c & d)**. Some intense states are present in PDOS for the O-2p orbitals of the isolated catechol molecule, as shown in **Fig. 7 (c)**. The intensity of these states decreases, and states near the Fermi level are depleted after the adsorption of the catechol molecule on the BP + Ti system, as seen in **Fig. 7 (d)**.

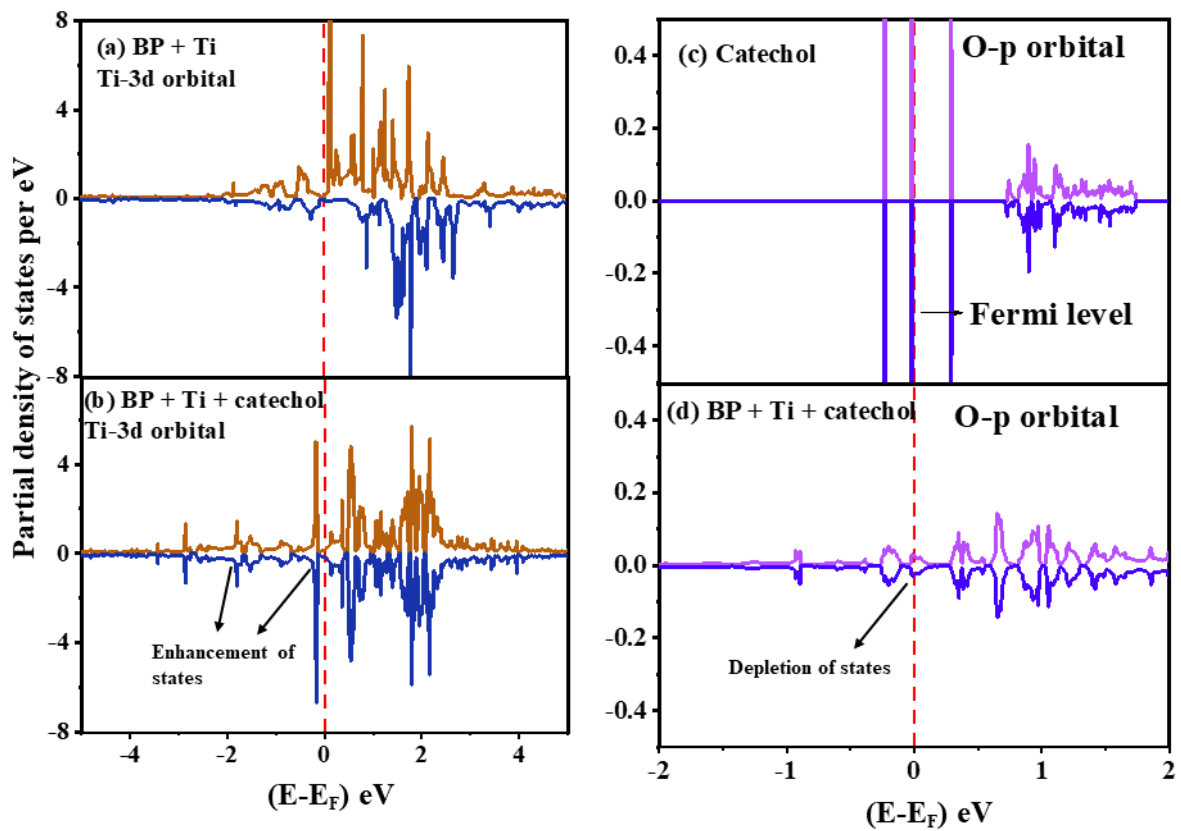

**Fig. 7** Partial density of states plots for (a) Ti-3d orbital of BP + Ti system (b) Ti-3d orbital of BP + Ti + catechol system (c) O-2p orbital of catechol molecule (d) O-2p orbital of BP + Ti + catechol system. The Fermi level is set to zero eV.

These reductions in states signify the transfer of charge from O-2p orbitals of the catechol molecule to Ti-3d orbitals, and this charge transfer is the reason for the strong binding of catechol to the BP + Ti system. Furthermore, we have plotted the PDOS for the Pd-4d orbitals and O-2p orbitals before and after the adsorption of the catechol molecule on the BP + Pd system, as presented in **Fig. S5 (a, b, c, & d)**. We may conclude from this PDOS plot that the charge is transferred from the O-2p orbitals of the catechol molecule to the Pd-4d orbitals of the BP + Pd system.

The sensitivity of a sensing substrate is related to the adsorption energy of the target molecule and the amount of charge transfer. Higher adsorption energy and a large number of charge transfers result in a strong chemical bond between the sensing substrate and the target molecule. From the Bader charge portioning, we found that the charge transfer amount from the catechol molecule is amplified to 0.10e for BP + Ti substrate, which was very less than 0.05e for pristine BP. For a pristine BP system, the catechol adsorption energy is very weak (-0.35 eV), leading to low sensitivity toward catechol. The strong interaction between BP + Ti with catechol leading to high adsorption energy -2.54 eV and significant charge transfer of 0.10e which can transform into a detectable signal and enhance sensitivity toward catechol.

The charge density difference plot between BP + Ti + catechol, BP + Ti, and pristine BP, i.e., $\Delta\rho = \rho_{BP+Ti+catechol} - \rho_{BP+Ti} - \rho_{BP}$, for the isovalue of 0.060e, is presented in **Fig. 5 (e & f)**. We have taken the B-G-R colour pattern to draw these plots where the red colour region on the catechol molecule represents the charge loss region, while the green colour regions near the Ti atom represent the charge gain regions, respectively. Hence, the charge is transferred from the catechol molecule to the metal Ti atom.

### 3.4 Practical viability of the system
#### 3.4.1 Diffusion energy barrier calculations

Metal clustering is a crucial issue for metal decorated nanomaterials that can reduce the sensing ability drastically. If the metal atom diffuses with ease from its adsorption site, then there is a huge possibility that the system may form the metal cluster. The experimental cohesive energy of the Ti and Pd atom is 4.85 eV/atom and 3.89 eV/atom, respectively [52]. The cohesive energy of the metal atoms is higher than the binding energies of -3.67 eV and -1.90 eV for Ti and Pd atoms, respectively. Thus, we have computed the diffusion energy barrier for the metal atoms by performing the climbing-image nudged elastic band (CI-NEB) calculations [53] to check the possibilities of the clustering of metal atoms into the BP sheet. We have calculated the energy barrier for the diffusion of the Ti atom from one hexagon site to the nearest hexagon site, while for the case of the Pd atom energy barrier is computed for the displacement along with one square site to the next square site as these are the most stable sites for the decoration of Ti and Pd atoms on BP sheet. The CI-NEB plots of the energy barrier versus diffusion coordinates for the movement of Ti and Pd atoms are presented in **Fig. 8(a & b)**.

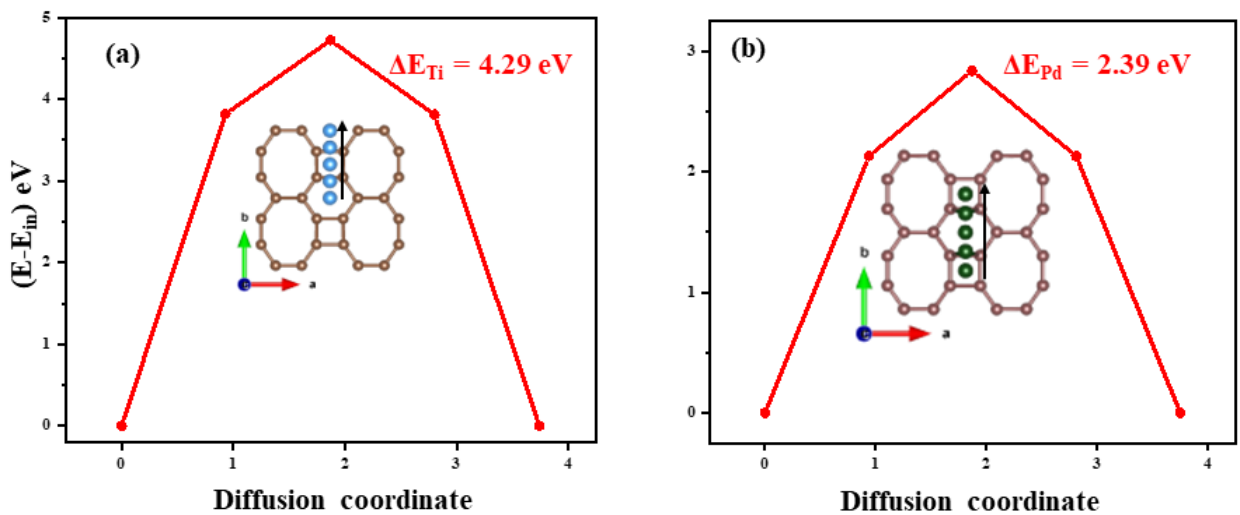

**Fig. 8 Diffusion energy barrier plots as a function of diffusion coordinates (a) for the movement of the Ti atom from above the centre of one hexagon site to another hexagon site of BP sheet (b) for the movement of the Pd atom from above the centre of one square site to another square site of BP sheet.**

The diffusion characteristics of doxorubicin on silica surfaces are studied by all-atom molecular dynamics simulation [54]. The authors have shown an analysis to calculate the diffusion coefficient from the classical Brownian motion equations. We have computed the diffusion energy barrier for the diffusion of metal atoms on the BP sheet by employing the standard CI-NEB method, which comes out to be 4.29 eV and 2.39 eV for the Ti and Pd atoms, respectively. There are various studies on the diffusion barrier of Ti and Pd atoms by applying CI-NEB method. For instance, Valencia et al. [55] have studied the adsorption of Ti atom onto graphene surface and calculated the diffusion energy barrier of 0.74 eV for the Ti atom. The diffusion barrier for the Ti atom on the MoS$_2$ surface is obtained at 5.96 eV [7]. The energy barrier for diffusion of the Ti atom over the graphene flake is reported to be 3.42 eV [56]. In divacancy defective graphene surface, the diffusion barrier of 5.82 eV exists for the Ti atom [57]. The diffusion barrier of 1.04 eV is found for the Pd atom on the graphyne surface [58]. The energy barrier of 1.90 eV for diffusion of the Pd atom is reported by Chen et al. [59]. Our calculated energy barriers are comparable with these previous reports and large enough to restrict the clustering me metal adatoms.

Moreover, the thermal energy of the metal atoms should also be smaller than the value of their diffusion energy barrier so that the system may remain prevented from clustering [60,61]. We have calculated the thermal energy of the metal atoms by using the relation, $E = \frac{3}{2} k_B T$, where E is the thermal energy, k$_B$ is the Boltzmann constant, and T is the temperature. We have taken T = 700 K; the maximum considered temperature for the recovery time calculation. The obtained thermal energy of the metal atoms is to be 0.09 eV. Thus, the computed energy barriers for the metal atoms are much higher than the thermal energy of the metal atoms, indicating that the metal atoms will remain attached to the BP sheet, and the metal decorated BP structures are prevented from metal-metal clustering.

### 3.4.2 Stability of the structure at room temperature

To investigate the stability of the metal-decorated BP system at room temperature, we have performed the *ab-initio* molecular dynamic (AIMD) simulations. The metal-decorated BP systems are placed in the microcanonical ensemble (NVE) for a time duration of 5ps, and the temperature is raised up to 300 K in a time step of 1fs. Then resultant system from the microcanonical ensemble is placed into the canonical ensemble (NVT) for a time duration of 5ps. The final MD replications of the BP + Ti and BP + Pd systems at 300 K are depicted in **Fig.9 (a & b)**, respectively. After the MD simulation, The Ti–C bond length changed to 2.52 Å, which was 2.17 Å before the MD simulation. Similarly, Pd–C bond length changes from 2.27 Å to 2.73 Å. So, we observe a minor modification in metal–C bond lengths; therefore, both systems are stable at room temperature.

Next, we have performed the AIMD simulations to check the integrity of the BP + metal (Pd, Ti) + catechol systems at ambient temperature. The final MD snapshots of BP + Ti + catechol and BP + Pd + catechol systems are presented in **Fig. 9 (c & d)**.

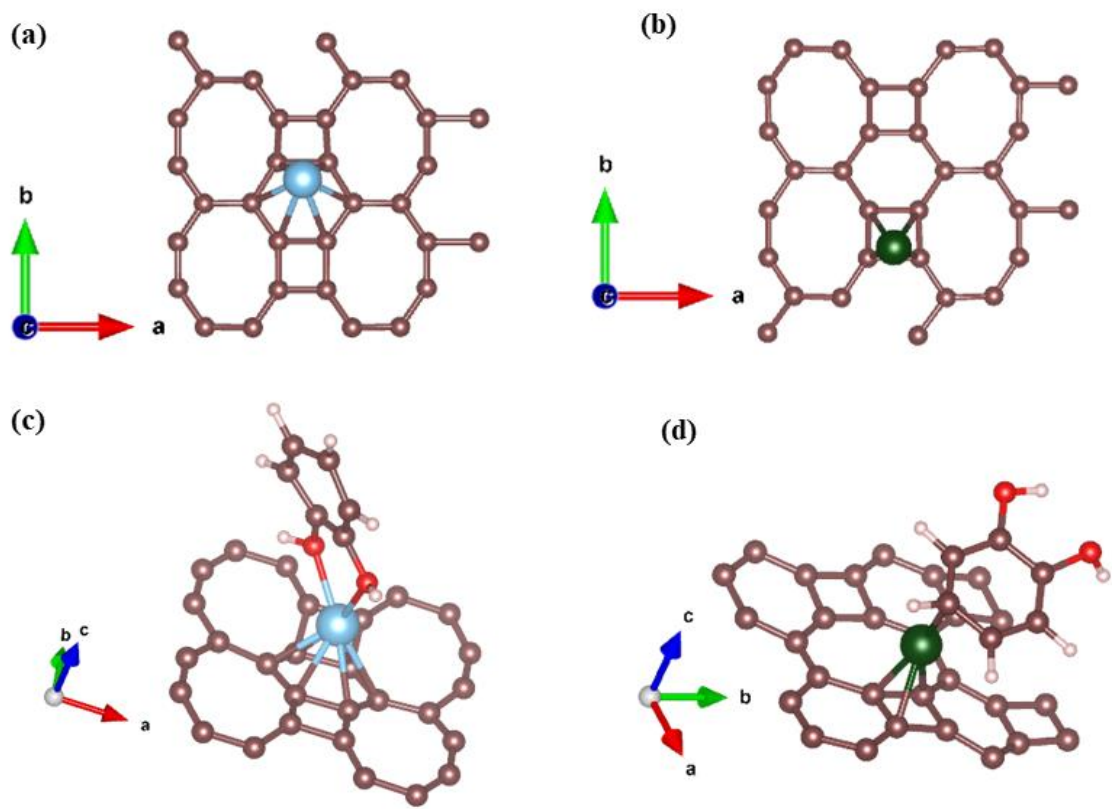

**Fig. 9 MD snapshots of (a) BP + Ti system (b) BP + Pd system (c) BP + Ti + catechol system (d) BP + Pd + catechol system, at room temperature and after putting the system in canonical ensemble for 5 ps.**

After the MD simulations, we got that both the structures are stable at room temperature. In the case of BP + Pd + Catechol, the orientation of catechol molecules changes with the rise in temperature, but the catechol molecule remains intact with the Pd decorated BP system even at room temperature. We have also plotted the bond length fluctuations between the carbon atom of biphenylene and the oxygen atom of catechol for the BP + catechol system (C-O), as presented in **Fig. 10 (a).** We found that the C-O bond length between catechol and pristine BP increases with the increase in temperature and becomes around 10 Å at 100 K. The C-O bond length remains more than 6 Å up to room temperature, indicating that the catechol molecule is desorbed from the pristine BP sheet even at the 100 K. Therefore, the pristine BP sheet is not suitable for catechol detection. Similarly, we have also plotted the Ti/Pd-O and Pd-C bond lengths for the catechol adsorbed on metal decorated BP systems, as shown in

**Fig. 10 (b & c)**. We have observed small thermal fluctuations in the Ti-O bond length, as shown in **Fig. 10 (b),** indicating that the catechol molecule remains intact to the Ti-decorated BP even at room temperature. In the case of Pd-decorated BP, we noticed that the orientation of the catechol molecule changes after around 200 K.

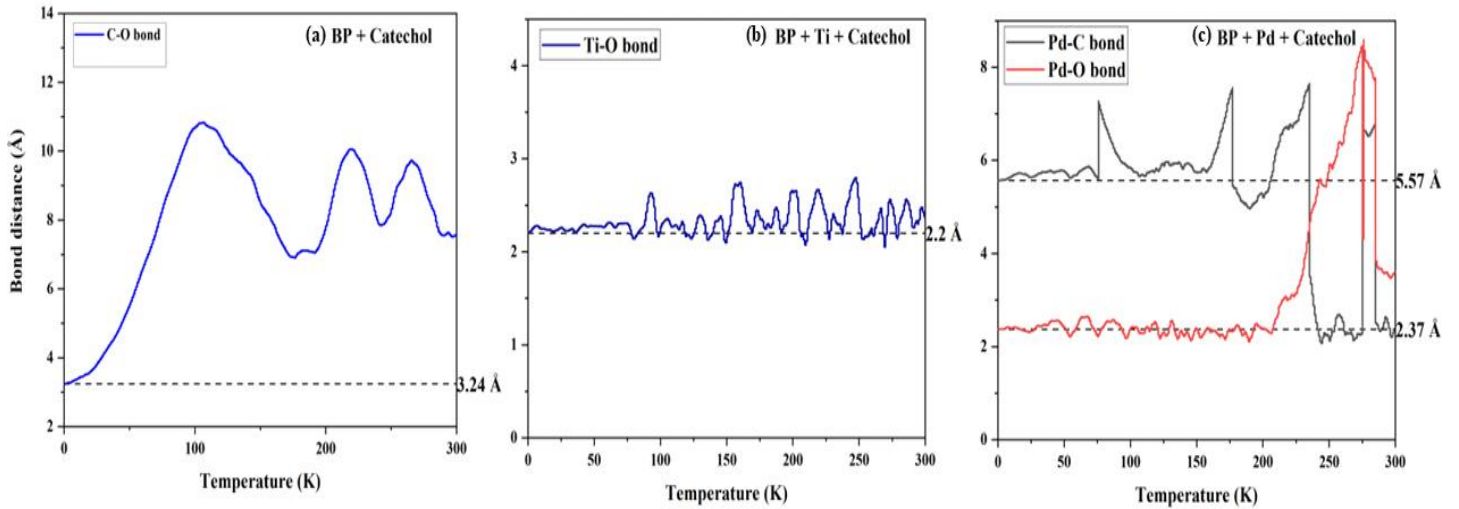

**Fig. 10 Bond length fluctuations with temperature (a) BP + Catechol system (b) BP + Ti + Catechol system (c) BP + Pd + catechol system.**

Therefore, we have plotted two bond lengths (Pd-O) and (Pd-C) atoms for the BP + Pd + Catechol system because when the orientation of catechol molecule changes the Pd-O bond length will increase, but the bond distance between Pd and the carbon atom of catechol will decrease, indicating that the catechol molecule is still attached to the Pd atom. The bond length fluctuations for the BP + Pd + Catechol system are displayed in **Fig. 10 (c).**

### 3.5 Recovery time calculations

The recovery time is the possible time duration in which an adsorbed biomolecule is desorbed from its host material. The recovery time necessarily consigns the reusability of the sensing material or device. For the practical feasibility of the system, recovery time should be less at ambient temperatures. The recovery time for the catechol molecule is calculated by using the transition state theory [24],

$$\tau = \vartheta^{-1} \exp\left(-\frac{E_b}{KT}\right) \quad (4)$$

where, $\tau$ is the recovery time, $\vartheta$ is the attempt frequency, $E_b$ is the binding energy, K is the Boltzmann constant, and T is the temperature. We have calculated the recovery time for BP + catechol, BP + Pd + catechol, and BP + Ti + catechol systems at different frequencies and temperatures [7,62]. To compare the recovery time of all systems precisely, namely, yellow light of frequency around $5.2\times10^{14}$ Hz and UV light of frequency $1\times10^{16}$ Hz are considered for the calculations. In addition to that, three different temperatures of 300 K, 500 K, and 700 K are considered. The calculated recovery times for yellow and UV light at all different temperatures are depicted in **Table 3.**

**Table 3. Recovery time calculations of different systems for visible (yellow) and UV lights at different temperatures.**

| System | Recovery time (sec) | | |
|---|---|---|---|
| | Yellow light 300 K | Yellow light 500 K | Yellow light 700 K |
| BP + catechol | $1.4\times10^{-9}$ | $6.0\times10^{-12}$ | $6.2\times10^{-13}$ |
| BP + Pd + catechol | $5.9\times10^{9}$ | 0.90 | $4.2\times10^{-11}$ |
| BP + Ti + catechol | $5.1\times10^{28}$ | $2.3\times10^{11}$ | $4.8\times10^{2}$ |
| | UV light 300 K | UV light 500 K | UV light 700 K |
| BP + catechol | $7.2\times10^{-11}$ | $3.3\times10^{-13}$ | $3.2\times10^{-14}$ |
| BP + Pd + catechol | $3.1\times10^{8}$ | $4.7\times10^{-2}$ | $2.2\times10^{-12}$ |
| BP + Ti + catechol | $2.6\times10^{27}$ | $1.2\times10^{10}$ | 25.2 |

Although, the recovery time of pristine BP is minimum, but the catechol adsorption energy on pristine BP is very less such that the catechol molecule gets desorbed even at 100 K. Hence, pristine BP is not feasible for detecting the catechol molecule. If we look at the case of yellow light frequency and 300 K temperature, the recovery time of BP + Pd + catechol and BP + Ti + catechol is quite high ($\sim10^{8}$); hence their reusability will not be feasible. When the temperature is raised to 500 K, the recovery time for BP + Pd + catechol system is 900 milli-second for the yellow light and 47 milli-second for UV light, which is suitable for reusability. But the BP + Ti + catechol system has a very high recovery time even at 500 K for both the frequencies and therefore, we further raised the temperature to 700 K, at which the recovery time for BP + Ti + catechol system decreases significantly so that the system may become feasible for reusability. The recovery time of BP + Ti + catechol at 700 K is 8 min for yellow light and 25.2 seconds for UV light.

From the simulation results, an essential advantage of using the BP sheet as a catechol sensing material is its suitable catechol adsorption energy. It is apparent from the recovery time calculation that the catechol adsorption energy should not be too high or too less. The previous study for catechol sensing on Sc embedded hGY system reports the catechol adsorption energy of -3.22 eV [22]. Although the large adsorption energy provides high sensitivity for catechol molecules, from a reusability point of view, it may lead to a very high recovery time and affect the reversibility of the system. The calculated recovery time for the $MoS_2$ + Pd system is $1.1 \times 10^2$ seconds at 500 K for UV light [7]. At the same condition, the recovery time for the BP + Pd system is 47 milliseconds. The sensing properties of various 2D materials for catechol detection are tabulated in Table 4.

**Table 4. Comparison between adsorption energy, charge transfer, and recovery time of different sensing materials for catechol detection.**

| Host material | Catechol adsorption energy (eV) | Maximum charge transfer | Recovery time at room temperature for visible light frequency (In seconds) |
|---|---|---|---|
| hGY + Sc [22] | -3.22 | 0.9e | - |
| $MoS_2$ + Ti [7] | -2.23 | 0.10e | $5.59 \times 10^{22}$ |
| $MoS_2$ + Pd [7] | -1.79 | - | $2.26 \times 10^{15}$ |
| $VSe_2$ + Pd [46] | -0.95 | 0.08e | 11.8 |
| **BP + Pd (present work)** | **-1.00** | **0.15e** | $\mathbf{5.9 \times 10^9}$ |

## 4 Conclusions

We have performed the first principles-based simulations to investigate the catechol detection properties of pristine and TM-decorated BP systems. We have selected Ag, Au, Pd, and Ti atoms for the decoration as they belong to different d-series and possess 1 and 2s valence electrons. The adsorption energy of catechol molecule on pristine BP sheet is -0.35 eV, which gets improved to -1.00 eV, and -2.54 eV with the decoration of Pd and Ti-atoms, respectively. By performing molecular dynamics simulations, we found that the catechol molecule gets desorbed at around 100 K from the pristine BP sheet but remains intact at room temperature when pristine BP is decorated with Pd and Ti atoms. From the Bader charge portioning, we found that O-2p orbitals of the catechol molecule transfer 0.1e charge to Ti-3d orbitals upon catechol adsorption on the Ti-decorated BP sheet. We calculated the diffusion energy barrier by performing the CI-NEB calculations to study the formation of metal–metal clusters in metal-decorated BP systems. The systems may be protected from metal-metal clustering with high diffusion energy barriers of 4.29 and 2.39 eV for Ti, and Pd atom, respectively. The stabilities of the BP + Pd + catechol and BP + Ti + catechol systems at room temperature are verified by performing the first-principles MD simulations. We calculated the recovery time at different frequencies and temperatures to investigate the reusability of the metal-decorated BP systems for catechol detection and found that the recovery time of the Pd decorated BP sheet is optimum at 500 K. In comparison, it is suitable for Ti-decorated BP sheet at an elevated temperature of 700 K.

## Conflicts of interest

There are no conflicts to declare.

## Author contribution statements

Vikram Mahamiya and Juhee Dewangan have equally contributed.

## Acknowledgement

VM and JD would like to acknowledge spacetime-2 supercomputing facility at IIT-Bombay for providing the computational facility. VM and JD would like to thank Mukesh Singh for his help. VM would like to acknowledge DST-INSPIRE for the fellowship. BC would like to thank Dr. T. Shakuntala and Dr. Nandini Garg for their support and encouragement. BC also acknowledges support from Dr. S. M. Yusuf and Dr. A. K. Mohanty.

# Supporting Information

# Remarkable enhancement in catechol sensing by the decoration of selective transition metals in biphenylene sheet: A systematic first-principles study

*Vikram Mahamiya[a], Juhee Dewangan [a], Alok Shukla[a], Brahmananda Chakraborty[b,c]*

[a]Indian Institute of Technology, 400076 Mumbai, India

[b]High Pressure and Synchrotron Radiation Physics Division, Bhabha Atomic Research Centre, Bombay, Mumbai 40085, India

[c]Homi Bhabha National Institute, Mumbai 400094, India

Email:mahamiyavikram@gmail.com, shukla@phy.iitb.ac.in, brahma@barc.gov.in

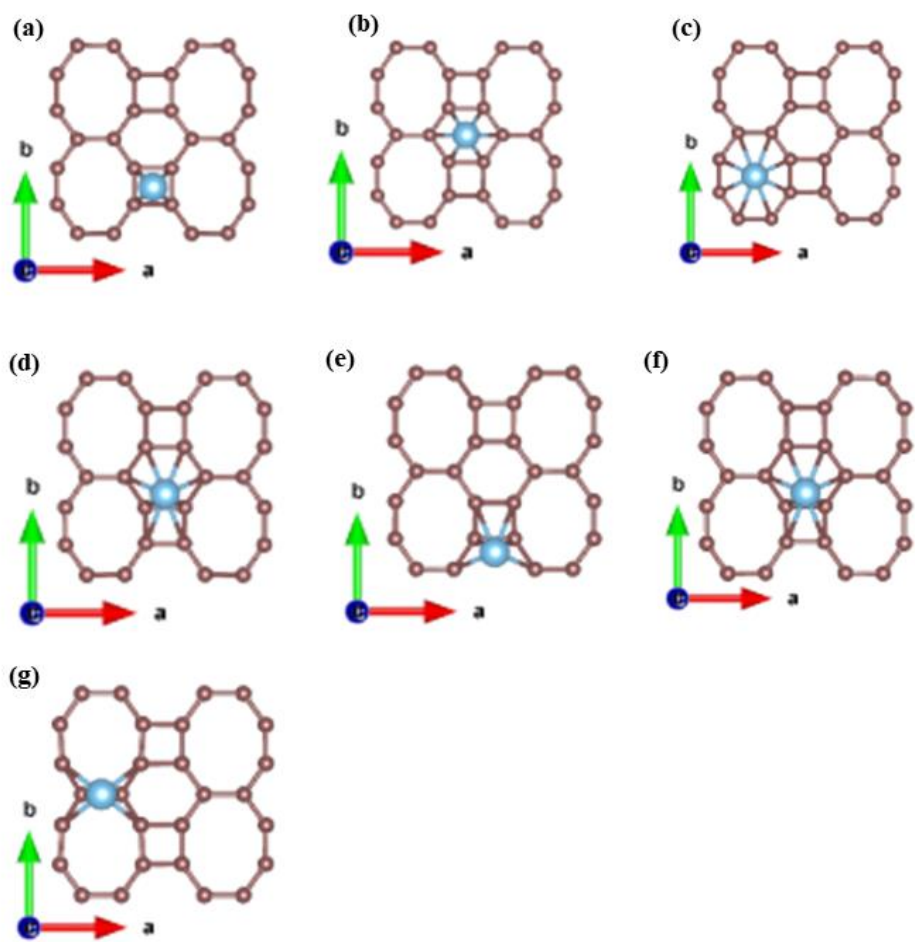

**Fig. S1** Optimized structure of Ti-decorated BP sheet, when Ti atom is decorated at various adsorption sites before the relaxation (a) Ti atom is placed above H1 site (b) Ti atom is placed above H2 site (c) Ti atom is placed above H3 site (d) Ti atom is placed above B1 site (e) Ti atom is placed above B2 site (f) Ti atom is placed above B3 site (g) Ti atom is placed above B4 site. The adsorption sites H1, H2, H3, B1, B2, B3, and B4 are described in Fig. 1(b). Here brown and blue colour balls represent carbon and Ti atom respectively.

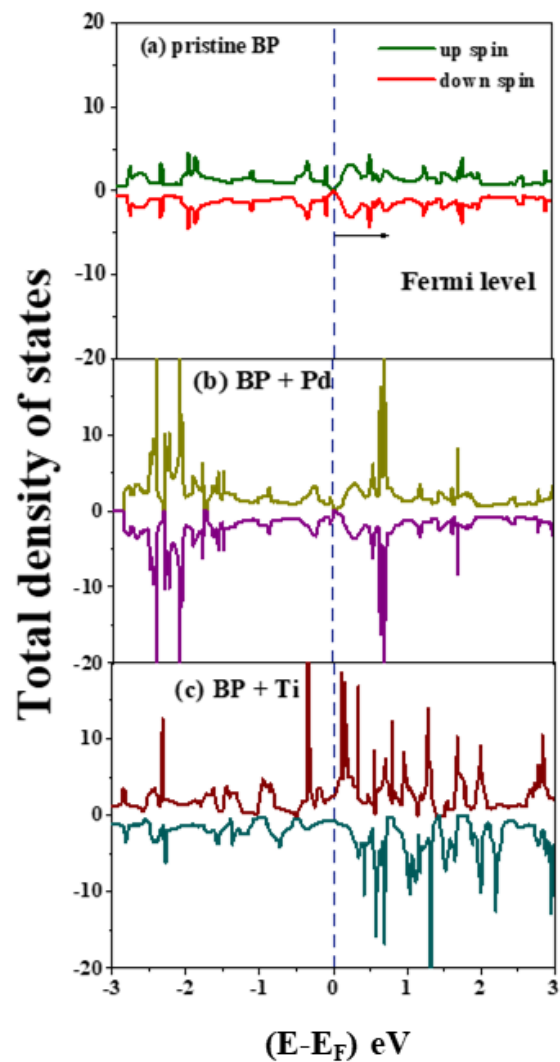

**Fig. S2** Total density of states plots for (a) pristine BP (b) BP + Pd (c) BP + Ti. The Fermi level is set to zero eV.

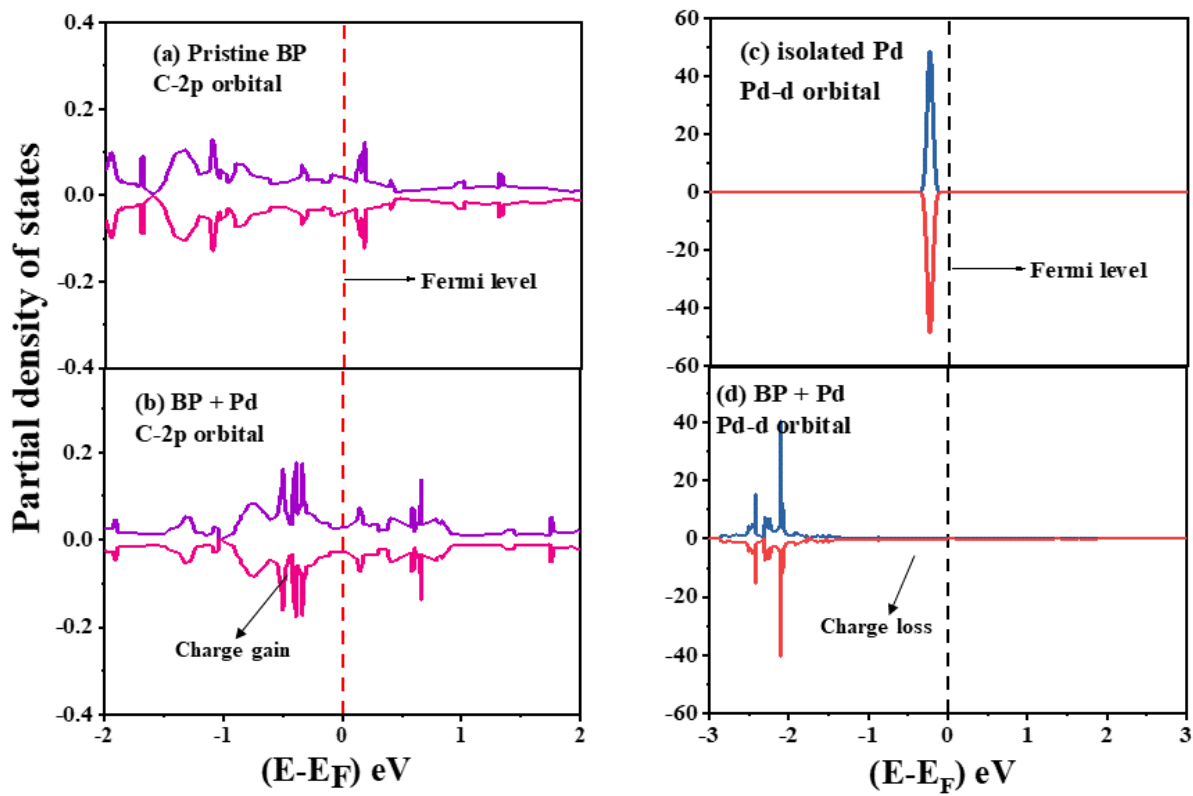

Fig. S3 PDOS plots for (a) C-2p orbital of pristine BP (b) C-2p orbital of BP + Pd (c) Pd-4d orbital of isolated Pd (d) Pd-4d orbital of BP + Pd. The Fermi level is set to zero eV.

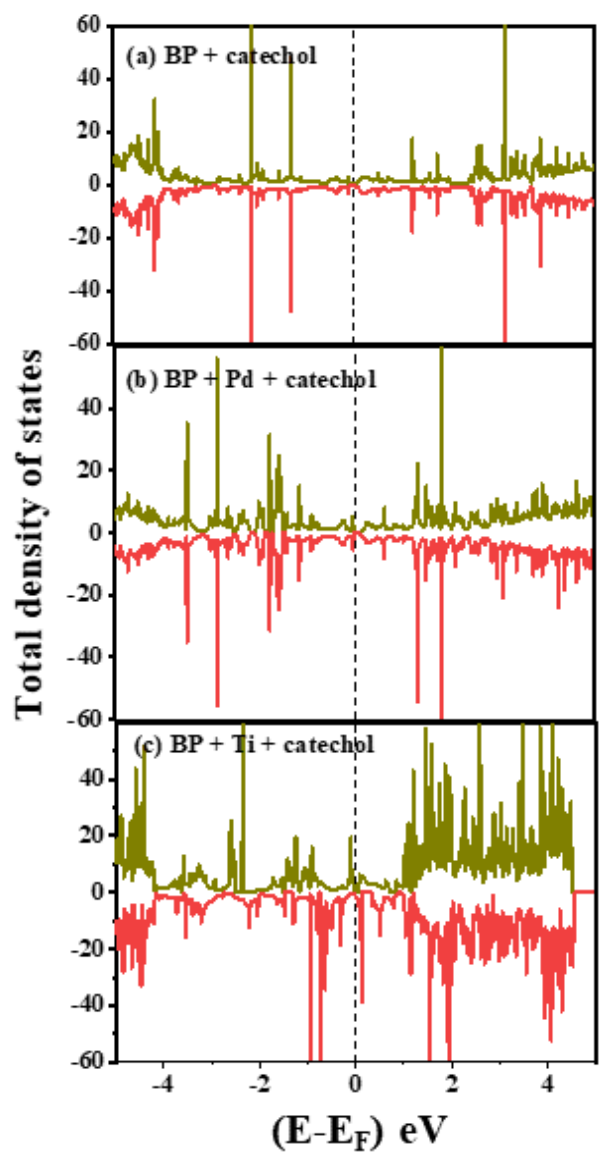

**Fig. S4** Total density of states plot for (a) BP + catechol system (b) BP + Pd + catechol system (c) BP + Ti + catechol system. The Fermi level is set to zero eV.

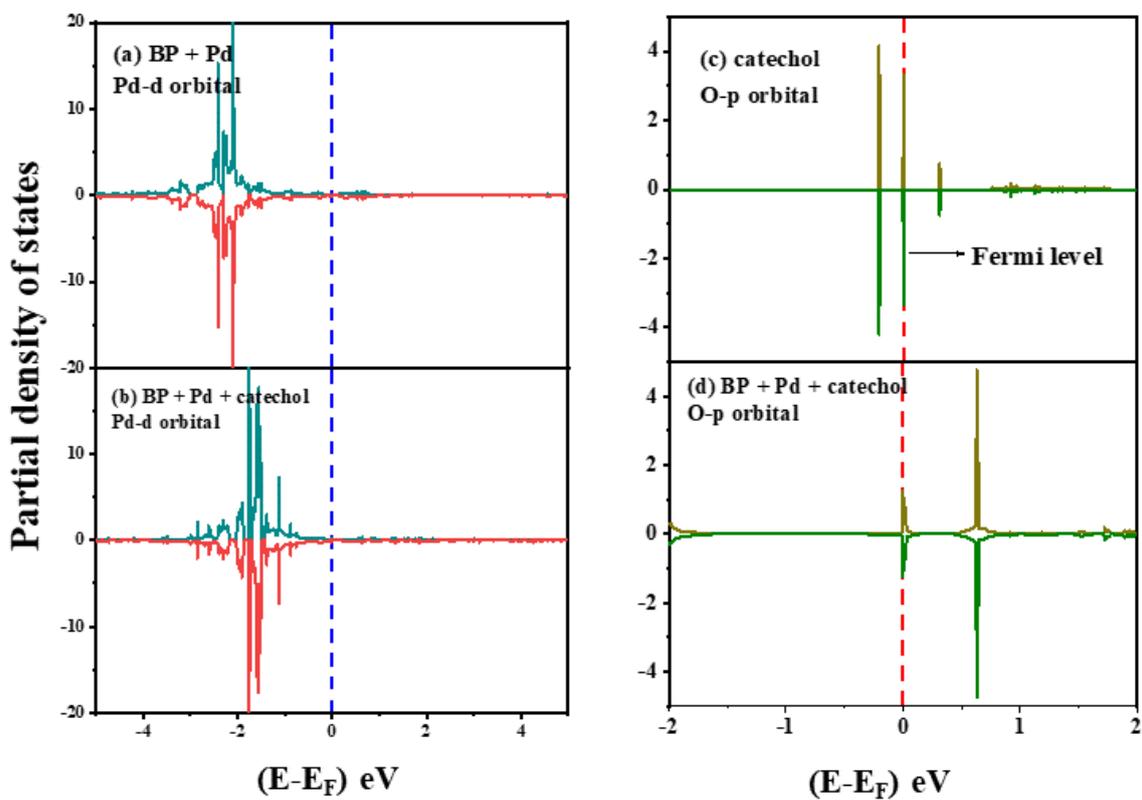

**Fig. S5** Partial density of states plot for (a) Pd-4d orbital of BP + Pd system (b) Pd-4d orbital of BP + Pd + catechol system (c) O-2p orbital of catechol molecule (d) O-2p orbital of BP + Pd+ catechol system. The Fermi level is set to zero eV.